\documentclass[conference, 10pt]{IEEEtran}
\IEEEoverridecommandlockouts 

\usepackage{cite}
\usepackage{amsmath,amssymb,amsfonts}

\usepackage{amsthm}

\usepackage[colorlinks,linkcolor=purple,citecolor=purple]{hyperref}
\usepackage{algorithm}
\usepackage{algpseudocode}
\usepackage{booktabs}
\usepackage{xcolor}
\algrenewcommand{\algorithmiccomment}[1]{\hfill\textcolor{blue}{$\triangleright$~#1}}

\usepackage[most]{tcolorbox} 
\usepackage{adjustbox} 
\usepackage[table]{xcolor}

\usepackage{caption}
\usepackage{graphicx}

\usepackage[normalem]{ulem}
\usepackage{tikz}

\newtheorem{theorem}{Theorem}
\newtheorem{corollary}{Corollary}
\newtheorem{proposition}{Proposition}
\newtheorem{proofnum}{Proof}

\newcommand{\sparseNone}{\tikz[baseline=-0.6ex]\draw[line width=0.4pt] (0,0) circle (0.8ex);}
\newcommand{\sparsePartial}{\tikz[baseline=-0.6ex]{
  \fill (0,0) -- (90:0.8ex) arc[start angle=90,end angle=270,radius=0.8ex] -- cycle;
  \draw[line width=0.4pt] (0,0) circle (0.8ex);
}}
\newcommand{\sparseFull}{\tikz[baseline=-0.6ex]\fill (0,0) circle (0.8ex);}

\definecolor{alg2color1}{HTML}{e3f1f3}

\definecolor{AtomPurple}{HTML}{7B35D5}
\definecolor{AtomBlue}{HTML}{4E79C6}
\definecolor{AtomCyan}{HTML}{31BFA9}
\definecolor{AtomXC}{HTML}{bf7a31}

\definecolor{MakoBlueDark}{HTML}{2F5DA8}
\definecolor{MakoBlue}{HTML}{4B7CC7}
\definecolor{MakoBlueLight}{HTML}{2FA7C8}
\definecolor{MakoRed}{HTML}{C73E3A}

\definecolor{FrontBoxBlue}{HTML}{F0F4F8}

\newtcolorbox{frontmatterbox}{
    enhanced,
    width=\textwidth,
    colback=FrontBoxBlue,
    colframe=FrontBoxBlue,
    boxrule=0pt,
    arc=12pt,
    outer arc=12pt,
    left=8mm,
    right=8mm,
    top=7mm,
    bottom=7mm,
    before skip=0pt,
    after skip=5mm
}

\def\BibTeX{{\rm B\kern-.05em{\sc i\kern-.025em b}\kern-.08em
    T\kern-.1667em\lower.7ex\hbox{E}\kern-.125emX}}

\begin{document}

\twocolumn[{%
\begin{@twocolumnfalse}
\thispagestyle{empty}

\noindent\includegraphics[height=0.48in,keepaspectratio]{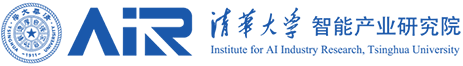}
\vspace{0.18in}

\begin{frontmatterbox}
{\raggedright
\fontsize{22.5}{25.5}\selectfont\bfseries
{\color{MakoBlueDark}M}{\color{MakoBlue}a}{\color{MakoBlueLight}ko}{\color{MakoRed}XC}:
Rearchitecting DFT Exchange--Correlation with
\underline{M}atrix-\underline{A}ligned and
\underline{K}nowledge-\underline{O}rganized Sparsity
\par}

\vspace{0.38cm}

{\raggedright
\fontsize{10.8}{13.2}\selectfont\bfseries
Haozhi Han$^{1,4}$,
Fusong Ju$^{2}$,
Jing Bai$^{3}$,
Ruge Zhang$^{5,4}$,
Xiang Zhao$^{5}$,
Liang Yuan$^{5}$,
Yunquan Zhang$^{5}$,
Ting Cao$^{4}$,
Yunxin Liu$^{4}$,
Yifeng Chen$^{1}$,
and Kun Li$^{4,\dagger}$
\par}

\vspace{0.22cm}

{\raggedright
\fontsize{9.7}{11.9}\selectfont\normalfont
$^{1}$School of Computer Science, Peking University, Beijing, China\\
$^{2}$Microsoft Research, Beijing, China\\
$^{3}$Microsoft Research, Redmond, USA\\
$^{4}$Institute for AI Industry Research (AIR), Tsinghua University, Beijing, China\\
$^{5}$Chinese Academy of Sciences, Beijing, China\\[2pt]
$^{\dagger}$Corresponding author.\\[2pt]
\textit{Part of this work was done at Microsoft Research and subsequently supported by PerfXLab Technologies.}
\par}

\vspace{0.44cm}

{\fontsize{9.8}{12.1}\selectfont\normalfont
Density Functional Theory (DFT) is indispensable for materials science and drug discovery, yet the exchange--correlation (XC) evaluation remains a major bottleneck due to its cubic scaling.
Although linear-scaling methods exploit electronic nearsightedness to reduce asymptotic complexity, they produce irregular sparse workloads that hide implicit sparsity and prevent efficient use of modern AI accelerators.
We present MakoXC, a modular matrix-aligned XC evaluation engine that rearchitects nearsightedness-induced sparsity into regular, accelerator-friendly computations. MakoXC co-designs three key techniques:
(1) Matrix-Aligned Cells reorganize nearsightedness-induced interactions into dense, accelerator-aligned data clusters;
(2) Sparsity-Guided Activation translates deeper implicit sparsity into numerically correct structured execution for practical linear scaling; and
(3) Kernel-Fused Pipeline consolidates fragmented workloads into a unified, compute-intensive execution path that fully unleashes accelerator throughput.
Extensive evaluations show that MakoXC achieves average speedups of 67.8$\times$ speedup over standard XC evaluation and 4.7$\times$ over state-of-the-art linear-scaling methods.
When integrated into a production-grade commercial DFT package, MakoXC scales XC evaluation to ubiquitin (1,231 atoms, def2-SVP) on 64 GPUs, enabling the end-to-end DFT calculation to complete in under five minutes.
By restructuring XC evaluation into a unified, structured computation, MakoXC demonstrates how scientific workloads can achieve genuine low complexity while maximizing parallel efficiency on AI accelerators.
\par}

\vspace{0.34cm}

{\raggedright
\fontsize{9.2}{11.2}\selectfont
\textbf{Email:}
\href{mailto:likun@air.tsinghua.edu.cn}{\textcolor{blue}{likun@air.tsinghua.edu.cn}}
\par}
\end{frontmatterbox}

\end{@twocolumnfalse}
}]

\section{Introduction}

{The rapid rise of artificial intelligence (AI) workloads has fundamentally reshaped the computing landscape~\cite{vaswani2017attention, wang2024comprehensive, wael2025accelerating}. Modern accelerators—particularly GPUs—are increasingly architected around AI-native operations, with specialized units such as Tensor Cores that execute fixed-shape matrix multiply–accumulate (MMA) tiles~\cite{markidis2018nvidia, schieffer2024rise, jouppi2017datacenter}. These units deliver peak throughput only when operands are dense, regular, and tile-aligned, thereby rewarding only computation that is uniform and highly structured~\cite{wang2019benchmarking, sun2022dissecting, sharma2021performance, wang2022benchmarking}.}

As a cornerstone of electronic-structure theory~\cite{kohn1996density, sherrill2010frontiers}, Density Functional Theory (DFT) exerts widespread impact on domains ranging from materials science~\cite{hafner2006toward, mattsson2004designing, neugebauer2013density} to drug discovery~\cite{ye2022applications, sabe2021current, tandon2019brief}. 
Its accuracy and efficiency, however, rely critically on the evaluation of the exchange–correlation (XC) contribution—the integration of energy densities and their derivatives over large, atom-centered grids~\cite{laikov1997fast, perdew2001jacob, martin2001integration, wheeler2010integration}. 
In practically relevant large-scale DFT calculations, numerical XC evaluation dominates the overall cost and becomes the main computational bottleneck.~\cite{stocks2025efficient, wang2024extending, laqua2018improved, burow2011linear}. 
With its conventional cubic-scaling complexity $\mathcal{O}(N^3)$, XC evaluation is both the most pressing challenge in practice and a compelling target for acceleration on modern hardware~\cite{stocks2025efficient, laqua2018improved, bartlett2005exchange}.

Attempts to accelerate XC evaluation on modern accelerators has evolved from dense formulations to linear-scaling methods.
As shown in Table~\ref{tb:intro},
dense formulations align naturally with hardware and achieve high utilization, but only by inflating computation and rigidifying the cubic complexity~\cite{yasuda2008accelerating, ufimtsev2009quantum, seritan2020terachem}. 
Linear-scaling methods exploit the electronic nearsightedness to lower the asymptotic complexity of XC evaluation from $\mathcal{O}(N^3)$ to near-linear $\mathcal{O}(N)$, enabling substantial performance gains at large scale~\cite{prodan2005nearsightedness, bader2008nearsightedness, prentice2020onetep, hernandez1996linear, scuseria1999linear}. 

However, the numerical manifestation of linear-scaling methods is highly \textbf{irregular sparsity}, which introduces two-level bottlenecks: 
{\textit{at the algorithmic level}}, 
the sparsity is exposed only in a fragmented form centered on individual atomic orbital (AO) values $\phi_\mu$, leaving the richer implicit sparsity induced by AO products $\phi_\mu\phi_\nu$ hidden and difficult to translate into usable algorithmic benefits~\cite{stocks2025efficient};
{\textit{at the hardware level}}, 
the irregular patterns lead to {fragmented execution} that misaligns with MMA tiling and fails to efficiently utilize modern AI accelerators~\cite{sun2022dissecting, manathunga2020parallel, dawson2018massively}.
As a result, linear-scaling XC evaluation still falls short of its theoretical promise in practice, leaving a substantial gap between algorithmic sparsity and efficient execution on modern AI accelerators.

\begin{table}[t]

\centering 
\captionsetup{font=small}
\caption{Comparison of Computational Structures and Execution Models of XC Evaluation in DenseXC~\cite{yasuda2008accelerating}, GPU4PySCF~\cite{wu2025enhancing}, and MakoXC.}

\label{tb:intro}

\begin{adjustbox}{width=1\linewidth}

\begin{tabular}{lcccc}
\toprule
XC & Scaling  & Matrix  & Sparsity & Execution \\
Method  & (Theo.)  & Structure & Exploitation & Feature  \\
\midrule
DenseXC & cubic & dense & \sparseNone  & monolithic   \\
GPU4PySCF & near-linear & irregular sparse & \sparsePartial & kernel-fragmented  \\
MakoXC & linear & aligned sparse & \sparseFull & pipeline-fused  \\
\bottomrule
\end{tabular}

\end{adjustbox}

\end{table}

In this paper, we propose \textbf{MakoXC}~\footnote{The name “\underline{Mako}XC” reflects its principle—\textit{\underline{M}atrix-\underline{A}ligned and \underline{K}nowledge-\underline{O}rganized Sparsity}—and evokes the Mako shark, renowned for its speed, symbolizing the pursuit of peak performance on modern AI accelerators.}, a modular XC evaluation system that rearchitects electronic-nearsightedness-induced sparsity into matrix-aligned structures, enabling implicit sparsity exploitation and highly parallel execution on Tensor Core GPUs.

The central insight is that sparsity, long viewed as an obstacle, can instead be restructured into {a bridge for algorithm–hardware co-design}, connecting algorithmic efficiency with hardware utilization.
Electronic nearsightedness induces global irregularity yet simultaneously creates local density. 
\underline{Mako}XC exploits this duality through the principle of {\underline{m}}atrix-{\underline{a}}ligned and {\underline{k}}nowledge-{\underline{o}}rganized sparsity: 
reorganizing irregular patterns into accelerator-aligned structures 
and unlocks implicit sparsity through domain knowledge and physical laws. 
Guided by this principle, MakoXC enables XC evaluation, for the first time, to unite practical linear complexity with high parallel efficiency on GPUs.

To realize this insight, MakoXC builds upon three tightly integrated techniques:

\textbf{Matrix-Aligned Cells (MACs).}  
MACs reorganize locally dense clusters hidden within global irregularity into fixed-shape, MMA-aligned cells by  
(1) clustering locality induced by electronic nearsightedness,  
(2) aligning computation with accelerator tiles, and  
(3) adapting to basis-set and grid resolution.  
This design transforms physically induced irregularity into accelerator-compatible units, preserving locality while exposing regular compute structures that enable subsequent implicit sparsity exploitation and efficient parallel execution.

\textbf{Sparsity-Guided Activation (SGA).}  
Building on MACs, SGA exposes the deeper implicit sparsity embedded in AO products and translates it into executable structured computation through a knowledge-organized two-stage activation scheme.
Going beyond explicit sparsity pruning, it first applies value-based filtering to identify numerically significant cells, and then uses AO product–density correspondence to activate only the density cells associated with significant AO products.
By fully exploiting both explicit and implicit sparsity induced by electronic nearsightedness, SGA enables structured computation with numerical correctness and practical linear complexity.

\textbf{Kernel-Fused Pipeline (KFP).}  
To translate structural advantages into real performance, 
KFP unifies XC evaluation into a fused on-chip pipeline that collapses the disjoint GEMM and reduction stages into a single tensor-core-centric dataflow, together with direct SRAM reuse and domain-knowledge-guided redundancy elimination, including symmetry folding and normcache gating.
This fusion transforms XC evaluation from a fragmented process dominated by frequent small-kernel launches, scheduling overheads, and memory traffic into a compute-intensive, high-throughput pipeline, fully unleashing Tensor Cores.

We evaluate MakoXC across three distinct molecular series and two widely used basis set configurations.
Our results show that MakoXC delivers average speedups of 67.8$\times$, 5.15$\times$ and 4.7$\times$ over DenseXC~\cite{yasuda2008accelerating}, GPU4PySCF~\cite{wu2025enhancing, li2025introducing, pu2025enhancing} and GauXC~\cite{williams2020efficient, williams2021achieving, 10.1063/5.0151070}, respectively, while exhibiting essentially perfect agreement with the theoretical linear-scaling trend.
Notably, when integrated into a mature commercial DFT package, MakoXC enables scalable XC evaluation for ubiquitin (1,231 atoms, def2-SVP) on 64 GPUs, allowing the end-to-end DFT calculation to complete in under five minutes and extending practical DFT simulations to larger molecular systems.

The main contributions of our paper include:

\begin{itemize}

    \item We introduce matrix-aligned cells, a hardware-native expression for electronic nearsightedness that unifies algorithmic sparsity with accelerator regularity.
    



    \item Built on the matrix-aligned foundation, we co-design sparsity-guided activation and kernel-fused pipeline to uncover implicit sparsity and unify fragmented XC workloads for practical linear scaling on modern accelerators.

    
    
    
    

    \item We implement these innovations as a modular XC engine compatible with mainstream DFT software and deployed in a production-grade commercial package, demonstrating broad integrability and industrial reliability.

    \item Beyond XC evaluation, this work advances HPC by showing that matrix-aligned formulation offers a viable design principle for translating irregular scientific sparsity into accelerator-native execution.
\end{itemize}

\section{Background and Challenges}

\subsection{Density Functional Theory and Exchange–Correlation Evaluation}

DFT describes the ground-state properties of an interacting electron system entirely in terms of its electron density~\cite{bartolotti1996introduction, kohn1996density, bickelhaupt2000kohn}. 
The ground-state energy can be written as a functional of $\rho$,
\begin{equation}
\small
E[\rho] = T_s[\rho] + E_\mathrm{ext}[\rho] + E_J[\rho] + E_\mathrm{xc}[\rho],
\end{equation}
where $T_s$, $E_\mathrm{ext}$, and $E_J$ denote the kinetic, external, and Coulomb terms, and $E_\mathrm{xc}$ is the XC contribution.
The XC {functional} $E_\mathrm{xc}[\rho]$ embodies the many-body effects of electron exchange and correlation, whose inherent complexity precludes any closed analytic representation~\cite{bartlett2005exchange, wheeler2010integration, koster2004efficient, balbas2001evaluation}.
As a result, $E_\mathrm{xc}$ must be approximated through density functionals evaluated by numerical integration~\cite{challacombe2000linear, laikov1997fast}.
To this end, progress along Jacob’s ladder systematically enriches XC {functional} dependence on the density and its derivatives, thereby improving accuracy, but simultaneously imposes greater computational demands, making $E_\mathrm{xc}$ the principal bottleneck in practical DFT calculations~\cite{perdew2001jacob}.

\textbf{Standard XC Evaluation.} In practice, XC contributions are evaluated numerically on atom-centered quadrature grids in three steps.  
First, the electron density is constructed on the grid:
\begin{equation}
\small
\rho(r_g) = \sum_{\mu\nu} \,D_{\mu\nu}\, \phi_\mu(r_g)\, \phi_\nu(r_g),
\label{eq:rho}
\end{equation}
where $D_{\mu\nu}$ is the density matrix and $\phi_\mu$ atomic orbital basis functions.  
{Second}, the XC functional is evaluated pointwise to produce energy densities and functional derivatives, a computationally light step~\cite{lehtola2018recent, williams2021achieving}.  
{Third}, the XC energy and Fock contributions are assembled by integrations:
\begin{equation}
\small
V^{\mathrm{XC}}_{\mu\nu} = \sum_g w_g\,\phi_\mu(r_g)\,v_{\mathrm{XC}}(r_g)\,\phi_\nu(r_g).
\end{equation}
Among these, Steps 1 and 3 dominate runtime, both requiring dense basis–grid integrations with complexity $O(N_{\mathrm{grid}} N^2)$ (basis size $N$ and number of grid points $N_{\mathrm{grid}}$). 
This cubic-like scaling rapidly becomes prohibitive for large systems~\cite{yasuda2008accelerating, ufimtsev2009quantum, seritan2020terachem}.

\textbf{Linear-Scaling XC Evaluation.} 
The key algorithmic innovation for overcoming this cost relies on electronic nearsightedness~\cite{prodan2005nearsightedness}: atomic orbitals (AOs) decay rapidly, so each grid point interacts only with nearby basis functions~\cite{junquera2001numerical, kudin2000linear, andzelm1992density}. Linear-scaling implementations exploit this property by introducing shell-dependent cutoff radii $r_{\text{cut},M}$, allowing distant AO contributions to be discarded~\cite{junquera2001numerical, bowler2012methods, sierka2003fast}. This reduces both density construction and potential assembly to near-linear scaling with system size~\cite{o2015linear}. However, the resulting AO-locality–driven sparsity is highly irregular in practice~\cite{williams2020efficient}. This irregular sparsity underpins all linear-scaling XC methods, yet simultaneously constitutes the central challenge we target in this work.


\subsection{Memory Hierarchy and Tensor Cores on GPU}

\begin{table}[b]
\centering 
\captionsetup{font=small}
\caption{Memory Hierarchy on A100 Tensor Core GPU.}
\label{table::memory_hierarchy}
\footnotesize
\begin{tabular}{l r r}
\toprule
\textbf{Memory Tier} & \textbf{Capacity} & \textbf{Latency (cycles)} \\
\midrule
Off-chip HBM (Global Memory) & 80 GiB / GPU & $\sim$290 \\
On-chip SRAM (Shared Memory) & 164 KiB / SM & $\sim$22 \\
On-chip SRAM (Registers, 32-bit) & 64 Ki / SM & $\sim$1 \\
\bottomrule
\end{tabular}
\end{table}

The NVIDIA GPU architecture is organized as a scalable array of multithreaded Streaming Multiprocessors (SMs), each hosting numerous CUDA Cores, specialized Tensor Cores, and a hierarchical memory system~\cite{jia2018dissecting}. 
Execution within an SM proceeds in warps of 32 threads, each equipped with its own register file. 
Threads within the same thread block share an on-chip SRAM space that offers low-latency, programmer-managed access, while all threads can access off-chip HBM, which provides larger capacity but significantly higher latency.
As summarized in Table~\ref{table::memory_hierarchy}, this hierarchy reflects the fundamental trade-off that increasing memory capacity comes at the cost of longer access latency~\cite{mei2016dissecting}.

Among the compute units, Tensor Cores are specialized for high-throughput matrix multiply–accumulate, defined as $D_{m \times n} = A_{m \times k} \times B_{k \times n} + C_{m \times n}$, delivering several-fold higher throughput than conventional CUDA Cores but restricted to fixed tile dimensions (e.g., $m=8$, $n=8$, $k=4$ in FP64 precision)~\cite{markidis2018nvidia}. 
While this design achieves exceptional efficiency for dense GEMM, the rigid tile geometry prevents effective utilization under the irregular and fragmented sparsity patterns of scientific workloads such as XC evaluation~\cite{sun2022dissecting}.

\subsection{Opportunity: Bridge Low Complexity to High Parallelism with Sparsity}
XC evaluation exposes a paradox: dense methods saturate hardware parallelism but rigidify cubic complexity, whereas linear-scaling methods reduce complexity yet yield irregular sparsity misaligned with accelerators. 
Crucially, this sparsity is not random but physically induced—and this distinction is key: while globally irregular, it exhibits local density and deeper implicit sparsity that can be systematically reorganized into regular, tile-aligned computational units and further mined for the implicit sparsity obscured by the native formulation. 
This rearchitecting not only aligns linear-scaling XC evaluation with modern GPUs, but also drives it from nominal near-linear scaling toward practical linear scaling. Sparsity, long regarded as an obstacle, thus becomes the key lever that bridges low algorithmic complexity with high accelerator parallelism.

\subsection{Challenges}

In this subsection, we outline three key challenges:

\textbf{Challenge \#1:}
\textit{How can nearsightedness-induced irregularity be reorganized into regular, accelerator-aligned units for efficient parallel execution?
}
In linear-scaling XC evaluation, the resulting workloads are fragmented and irregular, rendering them “anti-hardware” and unable to exploit modern accelerators. Existing remedies—such as enlarging batch sizes~\cite{williams2021achieving, manathunga2020parallel}, invoking CUTLASS device functions~\cite{stocks2025efficient}, or employing complex scheduling and memory control in libraries like DBCSR~\cite{sivkov2019dbcsr} and NTPoly~\cite{dawson2018massively}—only mask superficial overheads. The fundamental pathology of fragmented, irregular workloads persists.

\textbf{Challenge \#2:}
\textit{How can nearsightedness-induced sparsity be fully exploited as an algorithmic advantage to realize practical linear scaling?
}
Existing methods exploit only explicit sparsity from Grid–AO mapping at the level of individual basis functions~\cite{stocks2025efficient, besalu2011general}, while overlooking the richer, implicit sparsity inherent in AO products (e.g., $\phi_\mu(r_g)\phi_\nu(r_g)$). In principle, this implicit sparsity could greatly reduce computation and memory traffic. Yet doing so in a physically valid manner gives rise to highly irregular workloads, which have thus far precluded efficient mapping onto GPUs.
As a result, existing linear-scaling XC evaluation methods achieve linear scaling primarily in asymptotic theory, but have yet to realize true linear scaling in practical GPU execution.

\textbf{Challenge \#3:}
\textit{How can fragmented, memory-bound workloads be fused into unified pipelines to efficiently harness the throughput of modern AI accelerators?
}
Existing linear-scaling methods remain constrained by irregular workloads fragmented into numerous small kernel invocations.
Such fragmentation not only incurs substantial scheduling and launch overheads, but also produces workloads too small and uneven to sustain the high-throughput execution model on which modern AI accelerators rely.
Consequently, even though each individual kernel can often attain reasonably good implementation efficiency with vendor-optimized libraries, overall performance is still dominated by HBM traffic and launch overheads rather than computation. As a result, a large fraction of the accelerator’s throughput remains underutilized~\cite{stocks2025efficient}.

\section{System Design}

\begin{figure}[h]
    \centering
    \includegraphics[width=1\linewidth]{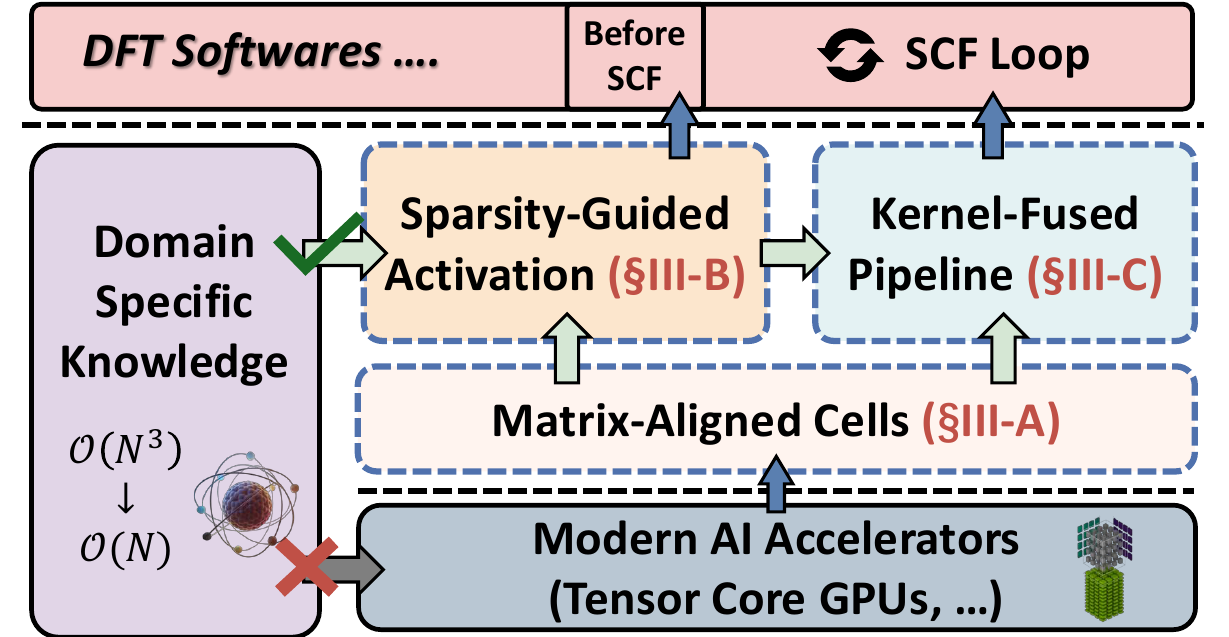}
    \caption{Overall Architecture of MakoXC.}
    \label{fig:intro-overall}
\end{figure}

As illustrated in Fig.~\ref{fig:intro-overall}, MakoXC is an algorithm–hardware co-designed system that maps domain-specific knowledge in XC evaluation onto the execution model of modern AI accelerators through three tightly coupled components.
Matrix-Aligned Cells (\S~\ref{3-1}) form the structural foundation, which reorganizes domain-specific knowledge-induced irregularity into regular, tile-aligned computational units for highly parallel execution on modern AI accelerators.
Built on this foundation,
Sparsity-Guided Activation (\S~\ref{3-2}) serves as the activation stage, which identifies the physically effective cells and computations through domain-specific knowledge, translating deeper implicit sparsity into executable structured sparsity embedded into mainstream DFT software before the SCF loop.
Built on the activated workload,
Kernel-Fused Pipeline (\S~\ref{3-3}) serves as the execution pipeline, which consolidates fragmented operations into a unified, high-throughput pipeline integrated into mainstream DFT software within the SCF loop.

\subsection{Matrix-Aligned Cells}
\label{3-1}

\begin{figure}[b]
    \centering
    \includegraphics[width=1\linewidth]{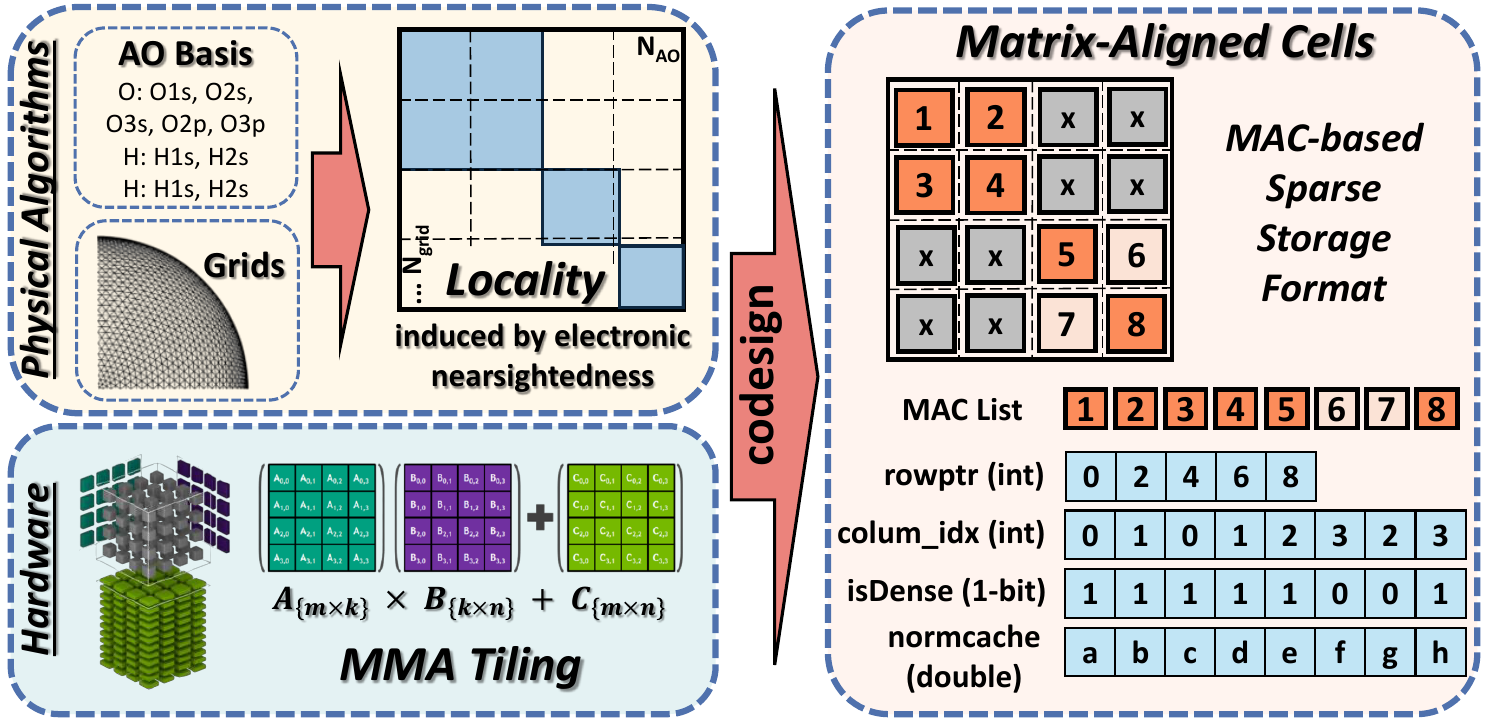}
    \caption{Matrix-Aligned Cells.}
    \label{fig:tech1-mac}
\end{figure}

In addressing the inherent tension between low algorithmic complexity and high parallel efficiency, the primary challenge lies in leveraging electronic nearsightedness—an essential source of near-linear algorithms—in a parallel-friendly manner. Traditional radius-based batching strategies can, in principle, achieve low complexity, but they induce highly irregular computations that misalign with the regular execution model of modern accelerators and incur substantial scheduling overhead.

To overcome this challenge, we propose Matrix-Aligned Cells (MAC), a mechanism that systematically maps the locality induced by electronic nearsightedness into regular computational structures. 
The core unit is the \textbf{MAC}, which encapsulates locality into dense micro-matrices tailored for efficient accelerator utilization. The design of MAC is guided by three key observations:
(1) {Local Clustering}: Owing to electronic nearsightedness, the Grid-AO matrix is globally sparse but locally dense, forming clusters of nonzeros due to localized basis overlaps and neighboring grid regions.
(2) {Alignment Optimality}: 
Modern accelerators (e.g., Tensor Core Units) execute matrix operations through fixed-size MMA tiles (e.g., $16 \times 16$). Aligning computation units to these tiles maximizes hardware efficiency.
(3) {Granularity Variability}: In DFT calculations, the choice of basis set and grid level directly influences the size of dense clusters.
Based on the above observations, the MAC embodies three essential properties:
(1) {Regularity}: Each MAC is a fixed-shape micro-matrix whose dimensions align with multiples of MMA tiles (e.g., $16 \times 16$, $32 \times 32$), ensuring compatibility with accelerator execution units.
(2) {Adaptivity}: The MAC size is not fixed but subject to a tradeoff: larger MACs yield more efficient matrix operations, but may partially compromise sparsity. At the beginning of a DFT calculation, the optimal MAC size is automatically determined based on the chosen quadrature grid and basis set, with the constraint that it must not exceed the available on-chip memory.
(3) {Sparsity-Awareness}: If a MAC exhibits extreme sparsity (e.g., for a $32 \times 32$ tile, fewer than 103 nonzero entries, i.e., $>90\%$ sparsity), it is stored in CSR format and marked as sparse by setting the corresponding entry in the $is\_dense$ array to 0.

While individual MACs encapsulate the local manifestations of electronic nearsightedness, leveraging this locality at scale requires a global perspective. 
We therefore design a \textbf{MAC-based sparse storage format}, which organizes local structures into a globally consistent layout, thereby enabling efficient parallel execution of XC evaluation.
As shown in Fig.~\ref{fig:tech1-mac}, the format employs a CSR-like index at the MAC granularity: $RowPtr$ (length $N_{\text{row}}+1$) stores the prefix offsets of MACs per row, so that the interval $[RowPtr[i], RowPtr[i+1])$ enumerates all MACs of row i; $ColumnIdx$ records the column index of each MAC. Each MAC is annotated with $isDense$, indicating whether it is stored densely or sparsely, and with a $normcache$, which holds a cached magnitude metric (in our case, the Frobenius norm) to support normcache gating {(\S~\ref{3-3})}. The actual data is stored in a contiguous MAC List, ensuring efficient access and parallel traversal.

\subsection{Sparsity-Guided Activation}
\label{3-2}

Building upon the MAC design, the next critical challenge is to fully exploit the electronic nearsightedness principle of linear-scaling DFT—maximizing the utilization of both explicit and implicit sparsity within the MAC paradigm.
To address this, we propose Sparsity-Guided Activation (SGA), a unified scheme operating through knowledge-organized sparsity. By enforcing knowledge-driven and physically rigorous filtering of MACs—activating only those with significant contributions while discarding all others as completely sparse—SGA transforms electronic nearsightedness into actionable computational efficiency, thereby pushing the practical complexity of XC evaluation ever closer to the theoretical linear-scaling limit.

\begin{figure}[b]
    \centering
    \includegraphics[width=1\linewidth]{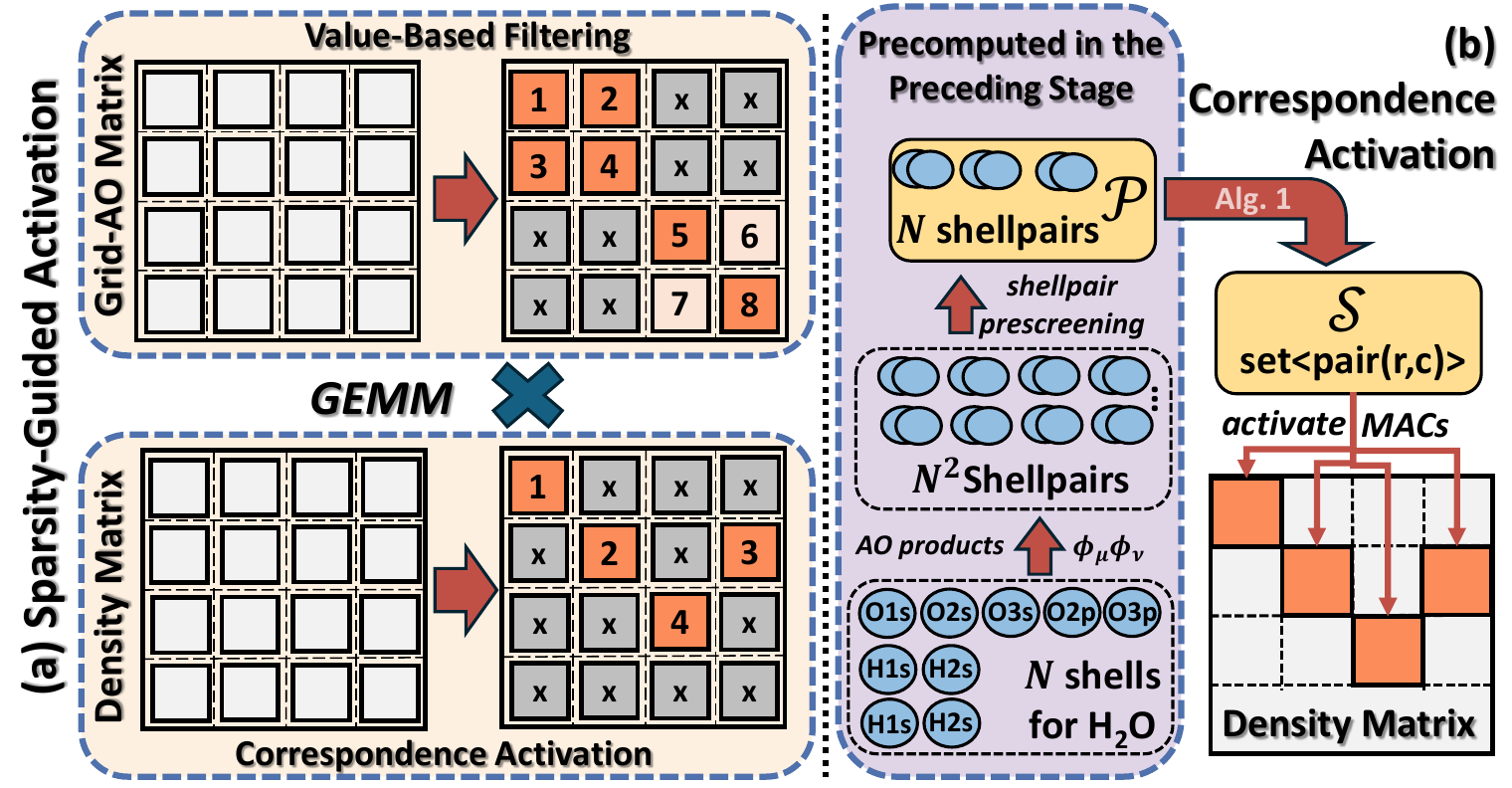}
    \caption{(a) Sparsity-Guided Activation. (b) Correspondence Activation for Exploiting Implicit Sparsity.}
    \label{fig:tech2-sga}
\end{figure}

SGA is founded on a key observation we term \textit{static sparsity}: the sparsity induced by electronic nearsightedness depends solely on the basis set and quadrature grid, and thus remains invariant throughout the SCF loop. Leveraging this property, SGA performs a one-time activation at the very beginning of the DFT computation, before the SCF loop. Thereafter, all subsequent XC evaluations are confined to the MACs activated in this initial SGA, enabling consistent acceleration across the entire SCF loop.
As shown in Fig.~\ref{fig:tech2-sga}(a), the design of SGA consists of two stages, namely Value-Based Filtering and Correspondence Activation.

\subsubsection{Value-Based Filtering} 
Within the theory of electronic nearsightedness, the exponential decay of Gaussian-type orbitals (GTOs) implies that their values diminish rapidly with increasing distance from atomic centers. Exploiting this property, SGA introduces a magnitude threshold $\varepsilon$ (default $10^{-12}$) on each AO basis function $\mu$ evaluated at grid points $r$. For any MAC in the Grid–AO matrix, if all elements fall below $\varepsilon$, the MAC is discarded outright; if more than 90\% of its elements fall below $\varepsilon$, the MAC is retained but stored in the CSR sparse format. The remaining MACs are deemed value-significant and are fully activated, preserved in the global MAC-based sparse storage format to sustain efficient downstream XC evaluation.

\subsubsection{Correspondence Activation}
For the Grid–AO matrix, SGA leverages Value-Based Filtering to uncover explicit sparsity.
However, this approach, akin to traditional radius-based batching strategies, only exploits explicit sparsity at the level of individual AO basis functions and thus falls short of revealing the deeper implicit sparsity inherent in the computation.

To address this, SGA introduces Correspondence Activation, which operates at the MAC level to further exploit the implicit sparsity embedded in the theory of electronic nearsightedness. This method is founded on the following key observations:
(1) \textbf{Richer combinatorial sparsity.} The evaluation of grid density involves AO products $\phi_\mu(r_g)\phi_\nu(r_g)$, whose product structure inherently contains numerous near-zero values. This implies abundant implicit sparsity that can be exploited to substantially reduce redundant memory accesses and computations.
(2) \textbf{One-to-one correspondence.} As can be seen from electron density formula (Equation~\ref{eq:rho}), each density matrix entry $D_{\mu\nu}$ has a strict one-to-one correspondence with the AO product $\phi_\mu(r_g)\phi_\nu(r_g)$.
(3) \textbf{Multiplicative property.} If an AO product is zero, the result of multiplying it with $D_{\mu\nu}$ is necessarily zero, even if $D_{\mu\nu}$ itself is nonzero. Consequently, such $D_{\mu\nu}$ entries can be excluded from storage and computation.
(4) \textbf{Static invariance.} Both the AO products $\phi_\mu(r_g)\phi_\nu(r_g)$ and the one-to-one correspondence remain invariant throughout the computation. 
Consequently, once a density matrix entry $D_{\mu\nu}$ is identified as having no contribution to the final result, this property holds throughout the entire DFT computation.
{Based on these observations, while the density matrix $D$ is dynamically updated and typically dense, its fixed correspondence with AO products induces a form of {static correspondence sparsity}: 
the mapping between $D_{\mu\nu}$ and AO products is invariant, and near-zero AO products render the corresponding $D_{\mu\nu}$ entries persistently sparse across the entire computation. 
This property not only exposes deeper implicit sparsity under electronic nearsightedness but also enables pre-activation at the beginning of the DFT computation by exploiting its static invariance.}

Therefore, Correspondence Activation requires only the prior identification of numerically significant (non-negligible) AO products and the localization of their corresponding positions within the density-matrix MACs, which are then activated; all remaining inactive MACs are deemed sparse and can be safely discarded.
In its ideal design, the numerical significance of an AO product is determined by directly evaluating the magnitude of $p_{\mu\nu}(r) = \phi_{\mu}(r)\phi_{\nu}(r)$. 
However, explicitly computing all AO products at every grid point is prohibitively expensive.
To address this challenge, we leverage the shellpair construction and its screening results—already precomputed in earlier stages of DFT—as an efficient surrogate for AO-product evaluation, thereby enabling the identification of MACs to be activated in the density matrix.

Importantly, the connection between shell-pair construction and the magnitude of AO products is not heuristic or empirical, but is grounded in rigorous mathematical principles. 
We next formalize why shell-pair construction serves as a principled surrogate for AO-product significance, ensuring a conservative screening criterion for Correspondence Activation—one that guarantees no false negatives.
\begin{theorem}[Gaussian Product Theorem]
For any two primitive Gaussian functions $\chi_\mu,\chi_\nu$ with exponents $\alpha_\mu,\alpha_\nu$ and centers $R_\mu,R_\nu$, the product can be rewritten as a single Gaussian:
\begin{equation}
    \chi_\mu(r)\chi_\nu(r) \;=\; P_{\mu\nu}(r)\, e^{-(\alpha_\mu+\alpha_\nu)\lVert r-R_c\rVert^2}\, e^{-\gamma \lVert R_\mu-R_\nu\rVert^2},
\end{equation}
where
\begin{equation}
\gamma_{\mu\nu} = \frac{\alpha_\mu \alpha_\nu}{\alpha_\mu+\alpha_\nu}, \qquad   
R_c = \frac{\alpha_\mu R_\mu+\alpha_\nu R_\nu}{\alpha_\mu+\alpha_\nu},
\end{equation}
and $P_{\mu\nu}$ is a polynomial whose degree equals the combined angular momenta of $\chi_\mu$ and $\chi_\nu$.
\end{theorem}
\begin{corollary}
The spatial magnitude of an AO product is concentrated near the line connecting the two shell centers, and its global amplitude is exponentially suppressed by the factor $e^{-\gamma \lVert R_\mu-R_\nu\rVert^2}$. Thus, when the shell centers are sufficiently far apart, $p_{\mu\nu}(r)$ is negligible for all grid points.
\end{corollary}
\begin{theorem}[Schwarz Bound]
The norm of an AO product satisfies
\begin{equation}
\lVert p_{\mu\nu}\rVert_2^2 = (\mu\nu|\mu\nu) \;\leq\; \sqrt{(\mu\mu|\mu\mu)} \,\sqrt{(\nu\nu|\nu\nu)} .
\end{equation}
\end{theorem}
\begin{corollary}
If the self-overlap integral of an AO is small, then its product with any other AO is necessarily small in the global sense. Hence, AO product significance can be strictly bounded by orbital self-integrals.
\end{corollary}
\begin{proposition}[Shellpair Significance Criterion]
Let shellpair $(P,Q)$ denote the set of all AO combinations from shells $P$ and $Q$. If either the shell-center distance $\lVert R_P-R_Q\rVert$ is sufficiently large or the Schwarz bound falls below a threshold $\varepsilon$, then for any $\mu\in P,\nu\in Q$,
\begin{equation}
    |p_{\mu\nu}(r)| < \varepsilon,\quad \forall r.
\end{equation}
\end{proposition}
\begin{proofnum}
By Theorem 1, the maximum AO product amplitude is controlled by the exponential decay term $e^{-\gamma \lVert R_P-R_Q\rVert^2}$; when the shell-center distance is large, this term tends to zero, ensuring insignificance. By Theorem 2, the global norm of the AO product does not exceed the Schwarz bound; if this bound is below $\varepsilon$, then the AO product is uniformly negligible. Hence, the proposition follows.
\hfill\qedsymbol
\end{proofnum}
Taken together, the significance of AO products is rigorously determined by inter-shell distances and orbital self-integrals—the very quantities exploited in shellpair construction and screening. Thus, the screened shellpair set constitutes a conservative surrogate for AO-product significance: while some negligible products may survive due to the use of upper bounds (over-activation), no genuinely significant contributions are ever discarded (no false negatives). This conservative guarantee preserves numerical correctness while dramatically reducing the cost of explicit AO-product evaluation, thereby providing both the theoretical rigor and practical efficiency that justify shellpair-based Correspondence Activation.

\begin{algorithm}[t]
\caption{Correspondence Activation on Density Matrix.}
\label{al:active-dm}
\begin{algorithmic}[1]

\small

\Require List of screened shellpairs $\mathcal{P}$, Dim of MACs $dim$
\Ensure Activated MACs in Density Matrix $\texttt{D}$

\State $\mathcal{S} \gets \emptyset$ \Comment{set of unique MAC indices $\mathcal{S} : (r, c)$}

\ForAll{$(x,y) \in \mathcal{P}$}
\For{$i \gets \texttt{sh2bf}[x] \ \textbf{to}\ \texttt{sh2bf}[x{+}1]-1$}
  \For{$j \gets \texttt{sh2bf}[y] \ \textbf{to}\ \texttt{sh2bf}[y{+}1]-1$}

    \State $r \gets \lfloor i/dim \rfloor$, $c \gets \lfloor j/dim \rfloor$
    \State $\mathcal{S} \gets \mathcal{S} \cup \{(r,c)\}, \{(c,r)\}$    \Comment{symmetric}
  \EndFor
\EndFor
\EndFor

\State $\texttt{D} \gets \textsc{Activation}(\mathcal{S})$

\end{algorithmic}
\end{algorithm}

As shown in Fig.~\ref{fig:tech2-sga}(b), SGA employs the shell pairs precomputed in the preceding stage to construct a set of unique MAC indices $\mathcal{S}$, which is automatically ordered and duplicate-free. 
Correspondence Activation is then realized by activating the MACs in the density matrix that correspond to the indices in $\mathcal{S}$.
The $\mathcal{S}$ construction process is detailed in Algorithm 1: for each shell pair, the procedure invokes the \texttt{sh2bf} function to locate all contracted Gaussian basis functions within the pair and determine their corresponding positions $(i,j)$ in the density matrix. These positions are then mapped to MAC indices $(r_b,c_b)$ and, by exploiting the symmetry of the density matrix, the resulting indices are inserted into $\mathcal{S}$. Finally, at Line 10, the effective MACs of the density matrix $D$ are activated based on the index set $\mathcal{S}$.
Correspondence Activation leverages the AO–density correspondence, rigorously activating only the significant MACs by exploiting shellpair screening and its strict equivalence to AO product significance.

In summary, SGA fully capitalizes on the sparsity induced by electronic nearsightedness: it first eliminates redundant MACs through value-based filtering, and then rigorously activates only the truly effective density-matrix MACs. This two-stage design uncovers both explicit and implicit of sparsity and, for the first time, translates implicit sparsity into a practically effective acceleration framework.

\subsection{Kernel-Fused Pipeline}
\label{3-3}

Before entering the SCF loop, MakoXC employs MAC to encode physical locality into regularized structural units and uses SGA to pre-filter and activate only those contributing to computation. 
To translate these structural advantages into hardware-level gains, it is imperative to design a kernel pipeline tailored to the MACs.

To this end, we propose the Kernel-Fused Pipeline (KFP)—a high-performance XC execution framework that systematically leverages the structural regularity of MACs to fuse critical operations, eliminate redundancy, and map execution efficiently onto modern accelerators, thereby unlocking fine-grained kernel-level optimizations and maximizing computational throughput.
KFP is motivated by three key observations:
(1) {HBM Bottleneck:} XC evaluation typically involves multiple GEMM and DOT operators; executing them independently makes HBM transfers the dominant performance bottleneck.
(2) {On-Chip Reuse:} Each MAC is sufficiently small to fit within the on-chip SRAM of modern GPUs, enabling cross-operator data reuse directly in fast SRAM.
(3) {Shape Alignment:} The consistent shape of MACs ensures that the input and output granularity across different operators is naturally aligned, eliminating additional data transformation overhead and making computation fusion virtually cost-free.
Building on these observations, KFP achieves operator-level fusion and on-chip reuse at the MAC granularity: by pipelining GEMM, DOT, and related operators within SRAM, it maximizes on-chip data reuse and drastically reduces HBM traffic. Meanwhile, the regularity of MAC shapes enables efficient operator fusion without requiring costly data reordering.

\begin{figure}[t]
    \centering
    \includegraphics[width=0.9\linewidth]{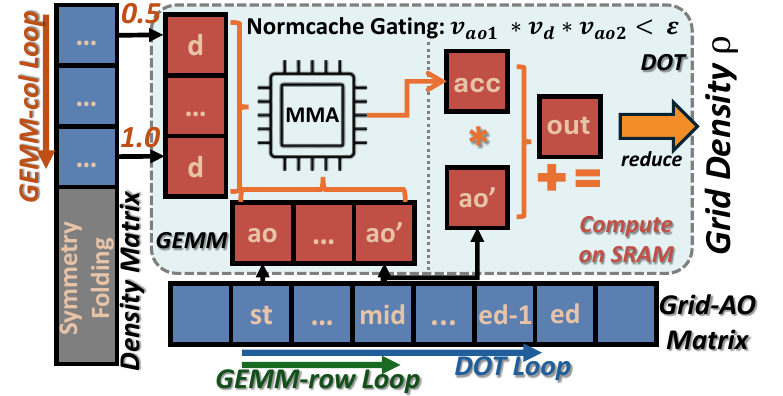}
    \caption{Kernel-Fused Pipeline.}
    \label{fig:tech3-kfp}
\end{figure}

\begin{algorithm}[t]
\caption{\textsc{ComputeRho} in Kernel-Fused Pipeline.}
\label{alg:tech3-kfp}
\begin{algorithmic}[1]
\small

\Require Sparse Grid-AO Matrix $AO$, Sparse Density Matrix $D$, threshold $\varepsilon$
\Ensure Grid density values $\rho$

\State $row = blockIdx.x;$
\State $start = AO.row\_ptr[row]$
\State $end = AO.row\_ptr[row + 1]$

\Statex \hspace*{-0.425em}%
\begin{tcolorbox}[colback=alg2color1, boxrule=0pt,
                  colframe=black!60,      
                  boxrule=0.8pt,        
                  left=0pt,right=0pt,top=0pt,bottom=0pt,
                  nobeforeafter,
                  width=\dimexpr\linewidth]

\State \textcolor[HTML]{205f9a}{\textit{\textbf{// DOT Loop}}}
\For{$mid \gets start$ \textbf{to} $end - 1$}

\State $v_{ao2} = AO.{norm}[mid]$
\State $col = AO.col\_index[mid]$
\State $D\_index = D.row\_ptr[col]$
\State $D\_end = D.row\_ptr[col + 1]$

\State $d_{row} = D.col\_index[D\_index]$

\Statex \hspace*{-0.275em}%
\begin{tcolorbox}[colback=red!15, boxrule=0pt,
                  colframe=red!60,      
                  boxrule=0.8pt,        
                  left=0pt,right=0pt,top=0pt,bottom=0pt,
                  nobeforeafter,
                  width=\dimexpr\linewidth + 0.275em]

\State \textcolor[HTML]{186b23}{\textit{\textbf{// GEMM-row Loop}}} \Comment{\textcolor{black}{\textbf{GEMM}}}
\For{ $A\_{index} \gets start$ \textbf{to} $mid$}  \Comment{\textcolor{blue}{Symmetry Folding}}
    \State $a_{col} = AO.col\_index[A\_{index}]$

    \State \textcolor[HTML]{c04f15}{\textit{\textbf{// GEMM-col Loop}}}
    \While{$d_{row} < a_{col}$ \textbf{and} $D\_index < D\_end$}
        \State $D\_index ++$
        \State $d_{row} = D.col\_index[D\_index]$
    \EndWhile

    \If{$a_{col} = d_{row}$}
        \State $v_{ao1} = AO.norm[A\_{index}]$
        \State $v_{d} = D.norm[D\_index]$

        \If{$v_{ao1} * v_{d} * v_{ao2} < \varepsilon$}
            \State \textbf{Continue} \Comment{\textcolor{blue}{Normcache Gating}}
        \EndIf

        \State $scale \gets (A\_index = mid)\ ?\ 0.5:\ 1.0$

        \State Load $ d = D.data[D\_index].T * scale$  
        \State Load $ ao = AO.data[A\_{index}]$

        \State \textbf{if} $ao.isDense \, \& \, d.isDense$ \textbf{then}
        \State \hspace{1.5em} \textbf{Tensor Core Compute:} $ acc = acc + ao \otimes d$
        \State \textbf{else}
        \State \hspace{1.5em} \textbf{CUDA Core Compute:} $ acc = acc + ao \otimes d$
        \State \textbf{end if}

    \EndIf
\EndFor

\end{tcolorbox}


\Statex \hspace*{-0.275em}%
\begin{tcolorbox}[colback=blue!15, boxrule=0pt,
                  colframe=blue!60,      
                  boxrule=0.8pt,        
                  left=0pt,right=0pt,top=0pt,bottom=0pt,
                  nobeforeafter,
                  width=\dimexpr\linewidth + 0.275em]
\State \hspace{-1.75em} $ao' \gets$ Load $AO.data[mid]$ \Comment{\textcolor{black}{\textbf{DOT}}}
\State \hspace{-1.75em} $out \gets out +  2 * ao' * acc$
\end{tcolorbox}

\EndFor

\end{tcolorbox}

\State \hspace*{-1.8em} StoreResults: $\rho \gets Reduce\_sum(out)$

\end{algorithmic}
\end{algorithm}

In this section, we take the kernel \textsc{ComputeRho} in XC evaluation as a representative example. 
The computation of grid density $\rho$ involves multiplying the Grid–AO matrix with the density matrix, followed by a dot product with the Grid–AO matrix on a per-grid-row basis. For clarity, we denote the former as $\Phi_1$ and the latter as $\Phi_2$, though they represent the same matrix: $\rho = (\Phi_{1} \times D) \cdot \Phi_{2}$.
Instead of issuing separate GEMM and DOT calls, KFP fuses them into a single pipeline to fully exploit on-chip data locality and minimize memory traffic. The overall workflow is shown in Fig.~\ref{fig:tech3-kfp}, with the corresponding pseudocode in Algorithm~\ref{alg:tech3-kfp}.

In \textsc{ComputeRho}, each thread block is assigned one row of $\Phi_2$ and processes all valid MACs within that row (after SGA pruning).
The first loop (Lines 5, DOT Loop) iterates over all valid MACs in the current row of $\Phi_2$, where the iterator $mid$ points to one such MAC, denoted as $ao'$.
Based on its position $(row, col)$ in $\Phi_2$, the corresponding MAC $acc$—with which MAC $ao'$ will be involved in a DOT computation—can be identified in the intermediate matrix $\Phi_1 D$, although this matrix is never explicitly materialized.
The position $(row, col)$ indicates that $acc$ is generated by multiplying all valid MACs in row $row$ of $\Phi_1$ with those corresponding in column $col$ of $D$. 
Accordingly, in Lines 12 and 15, the second and third loops (GEMM-row Loop \& GEMM-col Loop) traverse all valid MACs in row $row$ of $\Phi_1$ and column $col$ of $D$, respectively.
At Line 19, pairs of candidate MACs are checked for alignment in the matrix multiplication; once matched, the multiplication is executed on Tensor Cores or CUDA Cores.
If both $a$ and $d$ correspond to dense MACs, the computation is executed on Tensor Cores for maximum throughput, whereas sparse cases are executed on CUDA cores.
After the second and third loops, complete, $acc$ has already been computed and cached in SRAM. Without writing back to HBM, it can directly participate in the DOT operation with $ao'$ on SRAM. 
In Line 11-36, \textcolor{red!15}{\rule{0.75cm}{1em}} denotes GEMM operations executed on SRAM, while \textcolor{blue!15}{\rule{0.75cm}{1em}} denotes DOT operations executed on SRAM. Both are fused on-chip to maximize data reuse.
Finally, at Line 38, all MACs in row $row$ of $\Phi_2$ are reduced to obtain the corresponding density $\rho$.

The above description demonstrates how KFP transforms traditionally disjoint GEMM and DOT invocations into a unified on-chip pipeline, substantially reducing both memory traffic and kernel launch overhead. Crucially, by enforcing a global fusion perspective, KFP also exposes redundant computations and data flows that contribute negligibly to numerical accuracy—patterns that are largely invisible when operations are performed in isolation.

More importantly, the fused execution framework of KFP naturally creates opportunities to eliminate such inefficiencies: within a unified pipeline, operators, data streams, and storage paths are systematically organized, enabling patterns that were previously scattered and elusive to emerge.
Building on this insight, KFP further integrates two redundancy-elimination mechanisms:\\
\textbf{(1) Symmetry Folding:}
Because the density matrix $D$ is symmetric ($D_{\mu\nu}=D_{\nu\mu}$), it is unnecessary to load and compute the entire matrix; evaluating only its upper triangular part suffices. In Algorithm~\ref{alg:tech3-kfp} (Lines 12 and 25), KFP exploits this property by folding symmetric operations into one fused computation, thereby reducing memory loads and computation by nearly half. 
\\
\textbf{(2) Normcache Gating:}
Although MACs activated by SGA are themselves numerically significant, their contributions after GEMM+DOT computations may still be negligible (e.g., $v_{ao1} \cdot v_d \cdot v_{ao2} < \varepsilon$). 
Therefore, as shown in Algorithm~\ref{alg:tech3-kfp} (Line 22), KFP computes the product of the three MACs’ NormCaches before executing the triplet computation, compares it against the threshold $\varepsilon$, and filters out insignificant data streams on the fly. \\
By integrating Symmetry Folding and Normcache Gating into the fused pipeline, KFP retains only numerically significant computations while discarding redundant operations at the source, achieving both higher efficiency and stronger alignment with hardware execution.

KFP transforms XC evaluation from a memory-bound sequence of fragmented operators into a tightly fused on-chip pipeline, combining operator fusion with SRAM-level reuse to fully unleash accelerator performance. Beyond operator fusion, it further eliminates redundancy through symmetry folding and norm-cache gating, delivering a high-throughput execution pipeline for XC evaluation.

\begin{figure*}[t]
\centering
\includegraphics[width=1\linewidth]{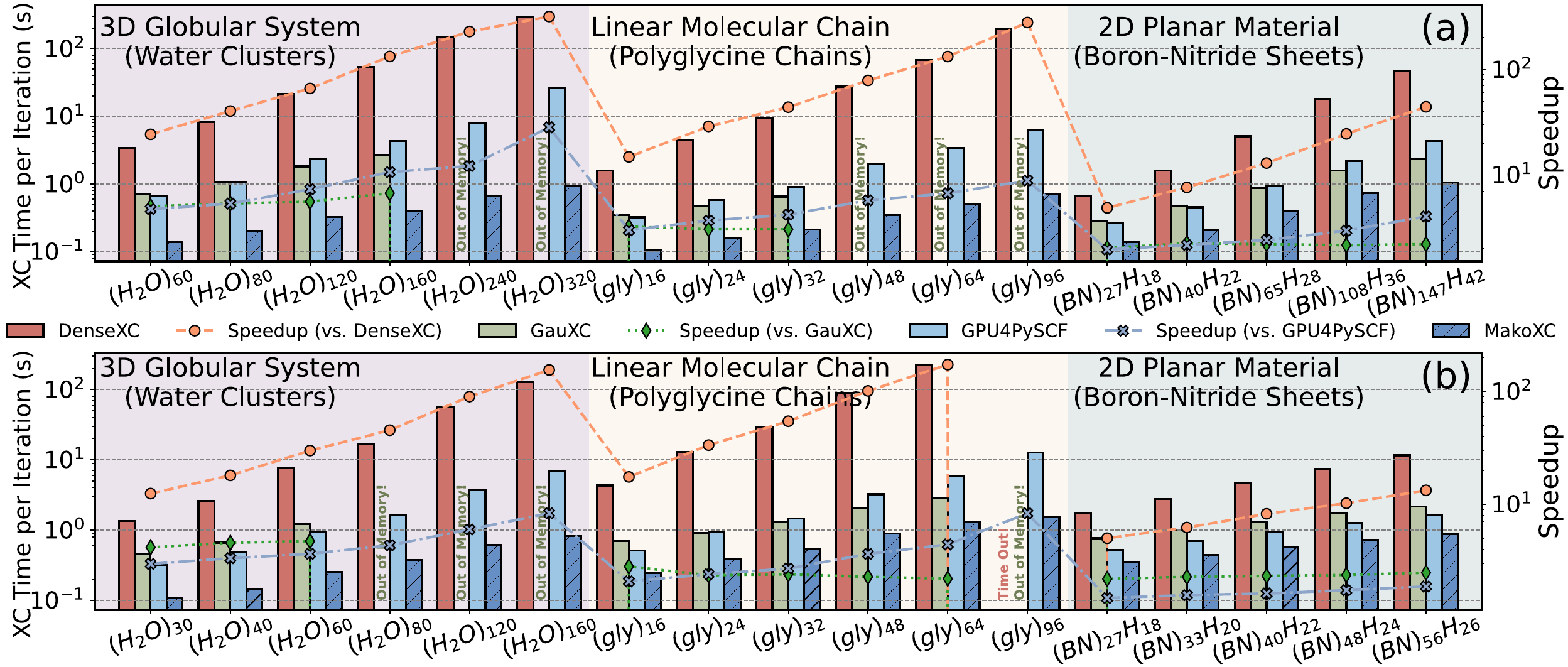}
\caption{Performance Comparison with State-of-the-art on a Single A100 GPU. (a) use def2-SVP, and (b) use def2-TZVP basis.}
\label{fig::overall-sota-comparison}
\end{figure*}

\section{Evaluation}
\subsection{Evaluation Setup}

\textbf{Hardware.} We evaluate performance on both single- and multi-GPU systems. The single-GPU setup consists of an AMD EPYC 7V13 CPU paired with an NVIDIA A100 Tensor Core GPU (Ampere architecture, Compute Capability 8.0). For multi-GPU evaluations, we employ a cluster where each node integrates 8 NVIDIA A100 GPUs and 96 AMD EPYC CPU cores, interconnected through 200 GB/s NVIDIA Mellanox HDR InfiniBand links.

\textbf{Metrics.} 
For standalone performance evaluation of MakoXC, we measure the average XC evaluation time within each SCF cycle~\cite{stocks2025efficient}. For end-to-end evaluation, we instead use the average iteration time of each SCF cycle~\cite{galvez2023high}.
The average is taken over the first ten iterations, excluding the initial one to eliminate initialization overhead, thereby ensuring consistent and fair timing across software packages with different convergence algorithms. 

\textbf{State-of-the-art.}
Since many DFT packages are closed-source, reproducing prior results remains difficult. To ensure a fair comparison, we evaluated several widely recognized open-source state-of-the-art packages, GauXC (v1.0)~\cite{williams2020efficient, williams2021achieving}, and GPU4PySCF (v1.4.1)~\cite{wu2025enhancing, li2025introducing, pu2025enhancing}. We also include DenseXC, our in-house implementation that ignores electronic nearsightedness, serving as a non-linear-scaling baseline.
Notably,  GauXC, which cannot independently perform SCF cycles, is integrated into our DFT framework and executed under the same interface to ensure fairness; however, its heavy reliance on global memory often leads to out-of-memory failures on larger molecular systems. 

\textbf{Dataset.} 
We have prepared three scalable classes of representative systems: 
(1) Polyglycine chains, a linear system scalable with chain length;
(2) Boron-nitride sheets, two-dimensional materials scalable with sheet size; and
(3) Water clusters, three-dimensional globular systems scalable with cluster size.
These scalable benchmarks provide a systematic basis for assessing MakoXC’s performance across different molecular dimensionalities and complexities.

\textbf{Aligned Parameter Settings.}
To ensure consistency and fairness in evaluating the performance of MakoXC, we adopted a unified set of parameter configurations across all experiments:
(1) Basis sets: We primarily tested the widely used def2-SVP and def2-TZVP basis sets, obtained from the Basis Set Exchange~\cite{exchange-basisset}.
(2) XC functional: All calculations employed the same B3LYP functional~\cite{becke1993new, stephens1994ab}.
(3) Integration grids: We adopted the same grid generation strategy as PySCF~\cite{sun2018pyscf} and GPU4PySCF~\cite{wu2025enhancing}. The grid level can be specified from 0 (very sparse) to 9 (very dense). 
Since different packages implement slightly different grid construction and grading schemes, we compared the corresponding grid levels by additionally reporting the number of grid points per atom. 
In all subsequent calculations, we fixed the grid level to 5, which corresponds to approximately 30,019 grid points per atom~\cite{lebedev1976quadratures, gill2003radial, mura1996improved}.
(4) Threshold ($\varepsilon$): For accuracy validation, we adopt the default threshold of $10^{-12}$, whereas for performance evaluation, we set the threshold to $10^{-10}$ to ensure consistency with other software packages~\cite{williams2020efficient, williams2021achieving}.

\subsection{Overall Performance}

\begin{figure}[t]
\centering
\includegraphics[width=1\linewidth]{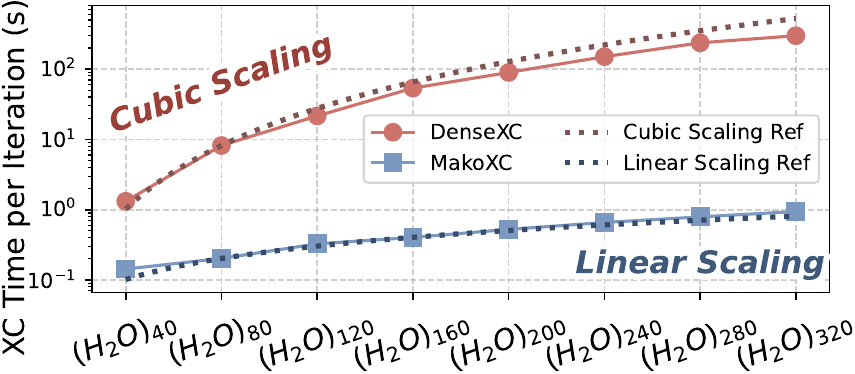}
\caption{Practical Linear Scaling Verification of MakoXC.}
\label{fig::linear-scaling-verification}
\end{figure}

\begin{figure}[t]
\centering
\includegraphics[width=1\linewidth]{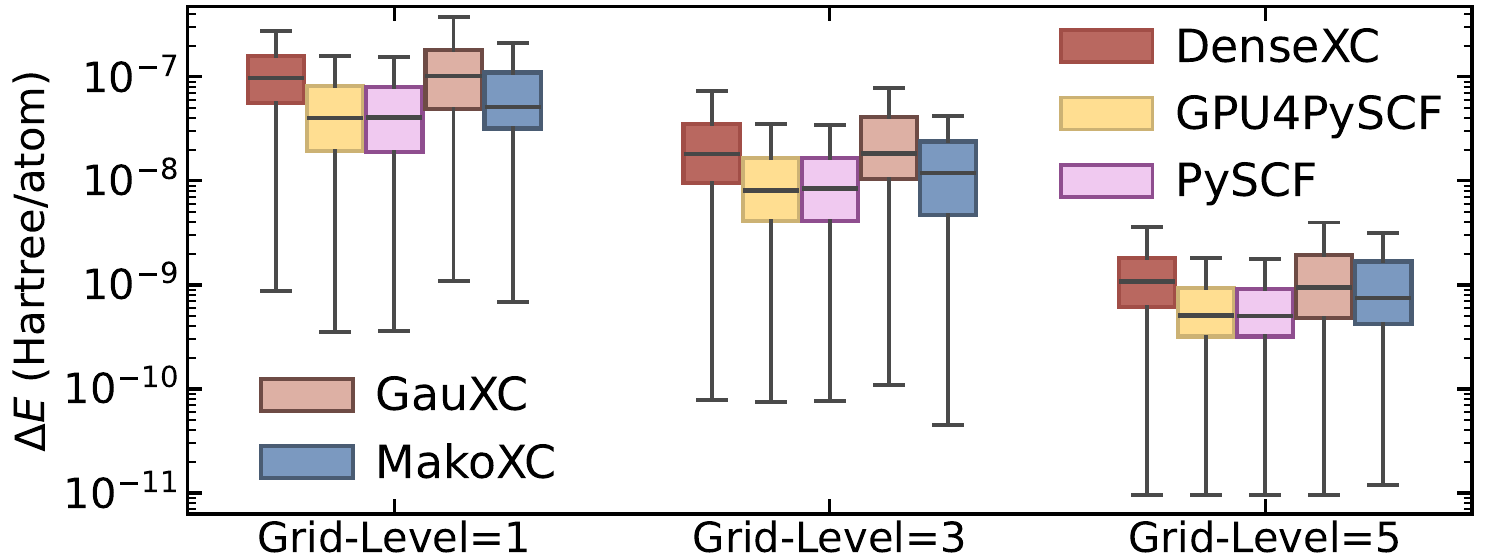}
\caption{Accuracy Validation of MakoXC.
Per-atom errors relative to Grid-Level=7 for all baseline across varying atomic grid densities with the threshold of \(10^{-12}\).}
\label{fig::accuray}
\end{figure}

\textbf{State-of-the-art Comparison.}
We evaluate the average XC evaluation time per SCF iteration and the corresponding speedup of MakoXC against all state-of-the-art software across diverse molecular systems and basis sets on a single NVIDIA A100 GPU. 
As shown in Fig.~\ref{fig::overall-sota-comparison}, MakoXC delivers substantial acceleration over all baselines. 
Under the def2-SVP basis set, MakoXC achieves an average speedup of 87.1× over DenseXC, 3.6× over GauXC, and 6.7× over GPU4PySCF. 
When moving to the larger def2-TZVP basis set, MakoXC continues to sustain substantial gains of 48.4× over DenseXC, 2.9× over GauXC, and 3.6× over GPU4PySCF, demonstrating that its performance advantage is sustained as system size and basis complexity increase.
These results demonstrate MakoXC’s efficiency, robustness across basis sets, and ability to fully exploit modern accelerators.

\textbf{Practical Linear Scaling Verification.}
Fig.~\ref{fig::linear-scaling-verification} plots the execution time of DenseXC and MakoXC as the system size increases proportionally, together with reference cubic- and linear-scaling curves. The results clearly show that MakoXC is the first method to realize practical linear-scaling XC evaluation as molecular size grows, turning what was previously only a theoretical complexity advantage into an achieved system-level behavior. To the best of our knowledge, MakoXC is also the first method to combine theoretical linear complexity with full alignment to modern hardware execution units, thereby fully unlocking their computational potential.

\textbf{Accuracy Validation.}
We evaluate the numerical accuracy of all XC evaluation baselines using per-atom energy differences (Hartree/atom), taking the grid-level=7 result from PySCF as the reference across varying atomic grid densities. As a widely used and well-established quantum chemistry package, PySCF provides a reliable numerical reference for accuracy evaluation.
To ensure a fair comparison, DenseXC, GPU4PySCF, PySCF, and GauXC use the same grid generation schemeand  screening threshold of \(10^{-12}\) as MakoXC. 
The energy error ranges of all methods are shown in Fig.~\ref{fig::accuray}.
Following the convention in quantum chemistry, XC energy errors within 1 mHartree/atom are regarded as chemically accurate~\cite{ju2024acceleration, doi:10.1021/ct100701w, tsuji2024dynamic}. 
Under this criterion, MakoXC exhibits excellent agreement with established software, thereby validating its correctness and reliability for practical quantum chemistry workloads.

\subsection{Ablation Study}

\begin{figure}[t]
\centering
\includegraphics[width=1\linewidth]{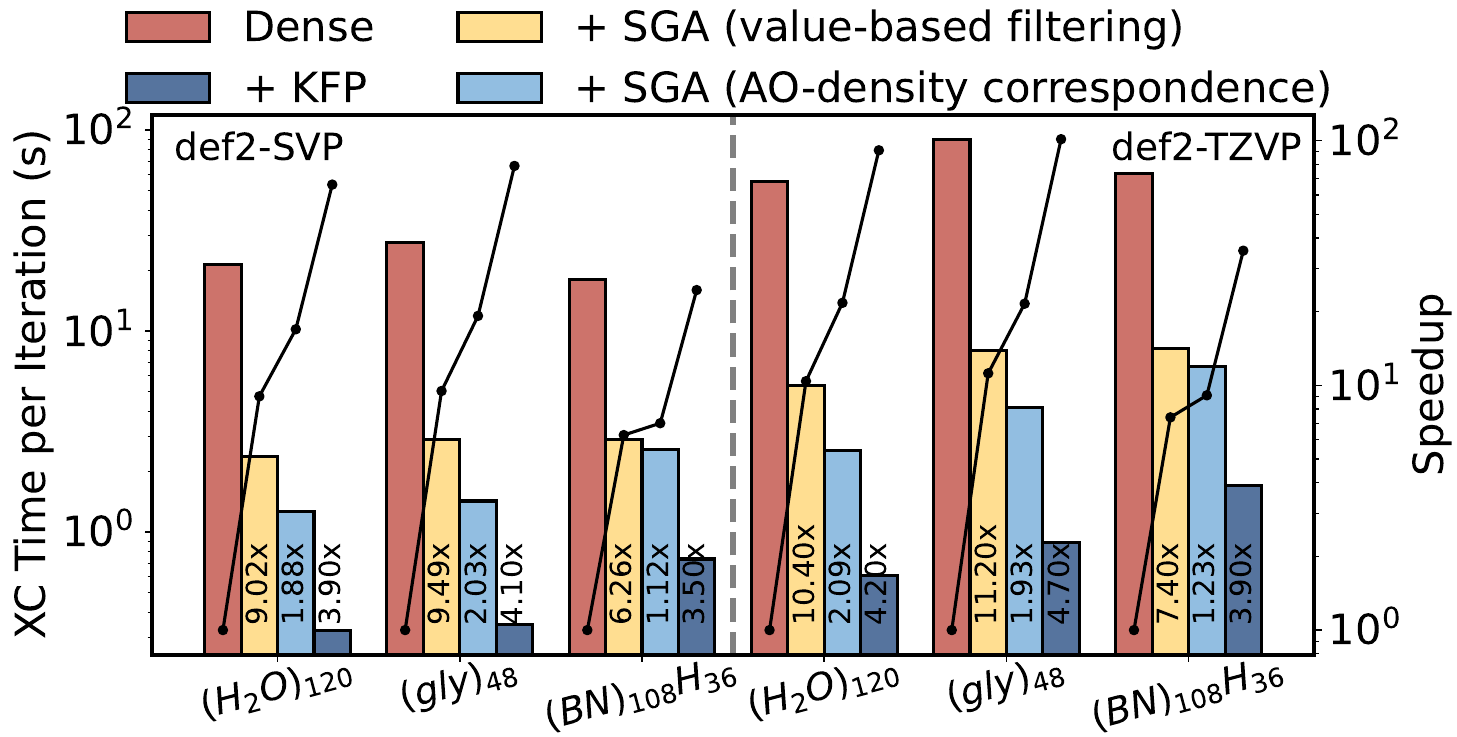}
\caption{Performance Breakdown of MakoXC.}
\label{fig::breakdown}
\end{figure}

\textbf{Performance Breakdown.}
Fig.~\ref{fig::breakdown} presents a performance breakdown of MakoXC across three representative molecular systems and multiple basis sets.
First, applying only value-based filtering—exploiting the sparsity of the Grid–AO matrix while ignoring that of the Density Matrix—delivers an average 9.0× speedup over the dense baseline. Second, employing the full SGA strategy, which additionally leverages Density Matrix sparsity, yields a further 1.71× improvement. Finally, when combined with the high-performance KFP implementation, MakoXC achieves an additional 4.1× acceleration on average.
These results demonstrate that each technique in MakoXC makes a substantial and complementary contribution to its overall performance improvement. 

\begin{figure}[t]
\centering
\includegraphics[width=1\linewidth]{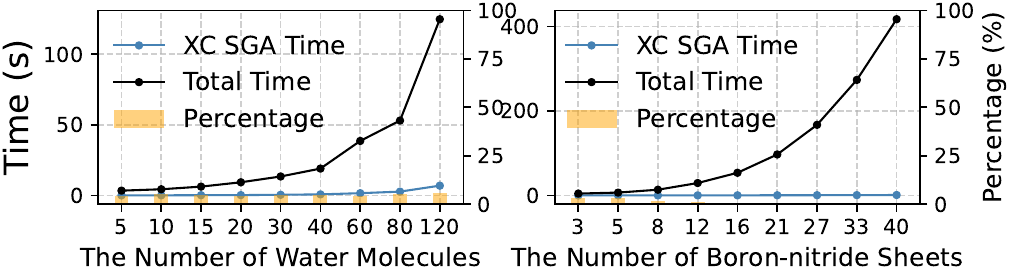}
\caption{Runtime Overhead of SGA before SCF Loop.}
\label{fig::sga-time}
\end{figure}

\textbf{Overhead Analysis of SGA.}
To ensure that the SGA procedure itself does not incur additional performance overhead, we measured its execution time. As shown in Fig.~\ref{fig::sga-time}, the XC SGA stage consistently accounts for less than 3\% of the total runtime across both globular molecular systems and boron–nitride sheets, with its relative cost further diminishing as system size increases. In contrast, the performance benefits of SGA are substantial, as evidenced by the performance breakdown results. Similar observations hold for all other tested molecular systems. These results decisively establish SGA as a highly effective and scalable strategy for XC evaluation.

\begin{figure}[t]
\centering
\includegraphics[width=1\linewidth]{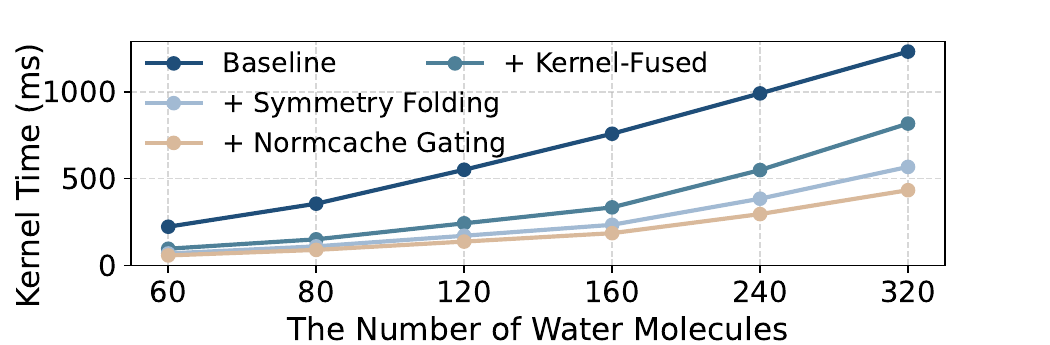}
\caption{Kernel Performance Improvements in MakoXC.}
\label{fig::kfp-kernel-time}
\end{figure}

\textbf{Breakdown of KFP.}
Fig.~\ref{fig::kfp-kernel-time} illustrates a breakdown analysis of these kernel techniques using the \textsc{ComputeRHO} kernel as a representative case, offering fine-grained insight into the performance improvements delivered by KFP. As shown, all three techniques consistently enhance kernel efficiency, with the fusion strategy yielding the most pronounced gains. Meanwhile, symmetry folding and normcache gating deliver increasing benefits as molecular system size grows. These results highlight the critical role of a high-performance kernel-fused pipeline design, which establishes a solid, hardware-aligned foundation for the overall MakoXC framework.

\subsection{End-to-End Evaluation and Scalability}

\begin{figure}[t]
\centering
\includegraphics[width=1\linewidth]{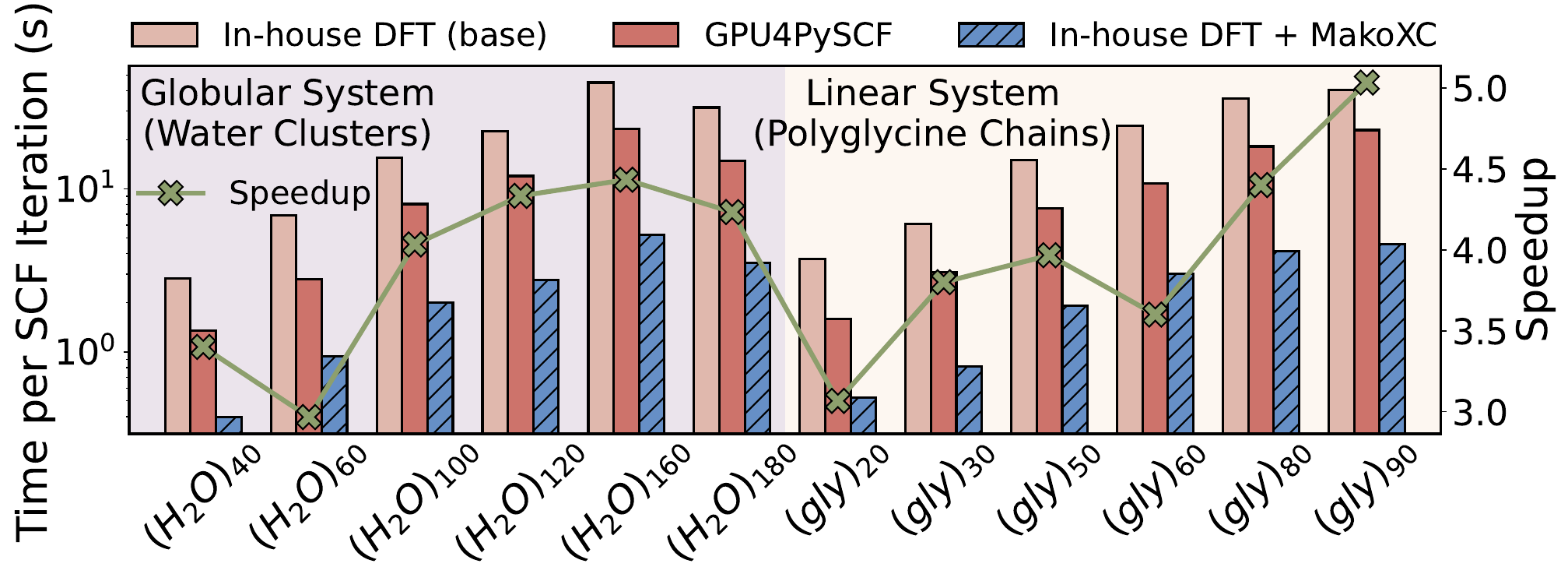}
\caption{End-to-End Performance Comparison. Speedup refers to the performance gain of the in-house DFT package with MakoXC over GPU4PySCF.}
\label{fig::end2end}
\end{figure}

\begin{figure}[t]
\centering
\includegraphics[width=1\linewidth]{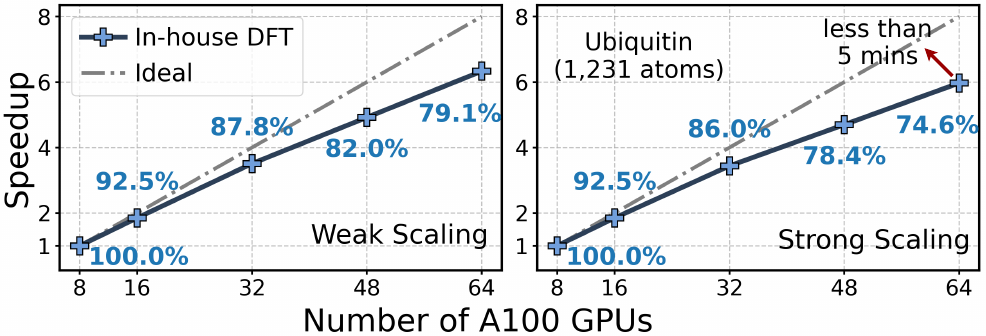}
\caption{Scalability of the In-house DFT Package with MakoXC.}
\label{fig::scalability}
\end{figure}

\textbf{End-to-End Evaluation.} We integrated MakoXC into our in-house DFT package, which provides a fully featured software stack together with highly optimized ERI kernels.
This integration leads to substantial end-to-end performance advantages over state-of-the-art implementations.
As shown in Fig.~\ref{fig::end2end}, the integrated system delivers an average 3.94$\times$ speedup over GPU4PySCF under the def2-SVP basis set.

\textbf{Scalability Evaluation.} 
Since GPU4PySCF provides limited support for multi-GPU execution, scalability testing was performed only on the integrated system.
XC integration is inherently amenable to multi-GPU parallelization, as each GPU can process a disjoint set of grid points with communication required only for the final reduction.
Using ubiquitin (1,231 atoms) with the def2-SVP basis set, we scale the system from 1 to 8 nodes (1–64 GPUs). As shown in Fig.~\ref{fig::scalability}, the system sustains over 74\% parallel efficiency even at 64 A100 GPUs across 8 nodes. 
At this scale, it completes a single-point energy calculation for ubiquitin in under five minutes, demonstrating both strong scalability and practical capability for large molecular simulations.

\section{Related Work}


DFT is a cornerstone of electronic structure calculations, but traditional plane-wave codes such as VASP~\cite{hacene2012accelerating, maniopoulou2012introducing, wende2019openmp, stegailov2019vasp, hafner2008ab} and Quantum ESPRESSO~\cite{barnes2017improved, giannozzi2009quantum, giannozzi2020quantum, scandolo2005first, carnimeo2023quantum} typically exhibit cubic scaling with system size, limiting their applicability to large-scale simulations~\cite{feng2024massively, jia2013analysis}. Linear-scaling DFT mitigates this limitation by exploiting electronic nearsightedness to approach \(\mathcal{O}(N)\) complexity~\cite{kohn1996density, burow2011linear, prentice2020onetep, hernandez1996linear, scuseria1999linear, junquera2001numerical, kudin2000linear, o2015linear}. Packages such as CP2K~\cite{hutter2014cp2k, kuhne2020cp2k, yokelson2022performance}, SIESTA~\cite{garcia2020siesta, artacho2008siesta, corsetti2014performance, sanchez2004computing}, and FHI-aims~\cite{blum2022fhi, abbott2025roadmap} employ localized basis functions and sparsity-aware algorithms to reduce computational complexity. 
However, the resulting computations often exhibit irregular sparsity and data-dependent execution patterns that are difficult to map efficiently onto modern GPUs~\cite{junquera2001numerical, lin2024ab, andzelm1992density, magalhaes2014gaussian, dunlap1990gaussian}. 
Recent work, Mako~\cite{10.1145/3712285.3759829}, demonstrates a matrix-centric alternative by rearchitecting irregular quantum-chemistry computations, particularly electron-repulsion integral (ERI) evaluation, into structured matrix-aligned kernels that can efficiently exploit modern AI accelerators. 
However, these techniques mainly target ERI evaluation, while exchange--correlation calculations exhibit distinct geometry-dependent, dynamically evolving sparsity that requires new matrix-aligned techniques.

Among the major components of DFT, exchange-correlation (XC) evaluation is a performance-critical stage because it lies on the critical path of every SCF iteration and repeatedly processes large numbers of grid points, basis functions, and density-dependent quantities~\cite{perdew2001jacob, stocks2025efficient, burow2011linear, bartlett2005exchange, koster2004efficient, balbas2001evaluation}. Although decades of GPU acceleration efforts—from early GGA kernel offloading to systems such as QUICK, GauXC, and GPU4PySCF—have substantially improved XC performance, the irregular sparsity in linear-scaling XC still causes fragmented workloads, irregular memory access, and load imbalance, leaving a large gap between its theoretical efficiency and realized performance on modern AI accelerators~\cite{yasuda2008accelerating, ufimtsev2009quantum, seritan2020terachem, manathunga2020parallel, manathunga2023quantum, williams2020efficient, williams2021achieving, wu2025enhancing, li2025introducing, pu2025enhancing}. This gap arises because the sparsity exposed by nearsightedness is numerically beneficial but computationally irregular, making it difficult to organize into the regular, high-throughput execution patterns favored by modern accelerator architectures.

\section{Conclusion}
MakoXC bridges linear-scaling XC evaluation with modern AI accelerators by uncovering implicit sparsity and reorganizing it into structured, matrix-aligned computation. It converts fragmented and irregular XC workloads into an efficient high-throughput execution path. This design enables practical linear scaling, significantly improves performance, and makes large-scale accurate DFT calculations practical.


\bibliographystyle{IEEEtran}
\bibliography{ref}

\end{document}